\documentclass[10pt,final,journal,twocolumn]{IEEEtran}

\usepackage[dvips]{graphicx}
\usepackage{float,makeidx,cite}
\usepackage{amssymb,amsmath,amsthm}

\providecommand{\abs}[1]{\lvert#1\rvert}
\newcommand{\ud}{\,\mathrm{d}}

\newtheorem{Propi1}{Proposition}

\newtheorem{Theoi1}{Theorem}
\newtheorem{Theoi2}[Theoi1]{Theorem}
\newtheorem{Theoi3}[Theoi1]{Theorem}

\newtheorem{Coro1}{Corollary}

\title{Rate Statistics in Cellular Downlink: A Per-User Analysis of Rateless Coded Transmission}
\author{Amogh Rajanna and Carl P. Dettmann}
\begin{document}
\maketitle
\begin{abstract}
In this letter, we focus on rateless coded adaptive transmission in a cellular downlink. Based on a stochastic geometry model for the locations of BSs, we evaluate the meta-distribution of rate, i.e., the distribution of rate conditioned on the point process. An accurate approximation to the distribution of per-user rate is proposed and clearly shown to provide a good match to the simulation results. We illustrate the gain in the per-user rate due to physical layer rateless codes relative to the fixed-rate adaptive modulation and coding.
\end{abstract}


\section{Introduction}
\label{sec:Intro}
One of the key technologies for 5G NR is the Raptor-like LDPC codes for the physical layer error correction\cite{ToRich}. It is anticipated that the future cellular standards/ technologies will evolve towards error correction schemes with more rateless-like properties. Modelling the locations of BSs and users by Poisson point processes (PPPs), it is shown in \cite{RHI,RDI} that an adaptive transmission based on physical layer rateless codes is very robust in terms of providing enhanced coverage and rate relative to the one based on fixed-rate coding and power control. In \cite{RDI}, the metrics used to compare the two  adaptive transmission schemes are the typical user coverage probability and rate. 
The typical user metrics are deterministic values and correspond to the spatial average of coverage probability and rate across the network, with the expectation taken w.r.t the point process. 

The meta-distribution of SIR is the distribution of coverage probability in the network conditioned on the point process\cite{MeDi_Pap}\cite{Wang}. It gives a fine-grained statistical probe into the location-dependent user performance across the network. In other words, the meta-distribution provides detailed information on the entire distribution of coverage probability rather than just the spatial average value across the network. In this letter, we focus on the coverage probability and rate conditioned on the point process. For the first time in the literature (to the best of our knowledge), this letter presents the location-dependent analysis of user performance in cellular downlink when physical layer rateless codes are used.

Using an accurate approximation for the distribution of bounded non-negative RVs, this letter quantifies the distribution of the per-user metrics, i.e., coverage probability and rate conditioned on the point process for rateless codes in cellular downlink. 
In \cite{MeDi_Pap}\cite{Wang}, the authors provide a location-dependent performance analysis when power control is used in cellular downlink. The inherent assumption is that fixed-rate coding is used. In such a setup, the only metric that needs to be characterized is the per-user coverage probability. The authors do not study the per-user rate in \cite{MeDi_Pap}\cite{Wang}. The per-user rate is obtained by scaling the per-user coverage probability by the fixed-rate of transmission. On the contrary, the main contribution of this letter is the characterization of per-user (location-dependent) rate in cellular downlink, which is very different from the metrics pursued in \cite{RHI,RDI,MeDi_Pap,Wang}. The simulation and analytical results in this letter show significant performance enhancements for the per-user rate in cellular downlink due to rateless codes relative to adaptive modulation and coding, and fixed-rate coding with power control. 
\section{System Model}
\label{sys_mod}
We consider a single tier cellular downlink in which the locations of BSs are modeled by a PPP $\Phi \triangleq \Phi_b \cup \{o\}$, where $\Phi_b=\{X_i\},~i=1,2,\cdots$ is a homogeneous PPP of intensity $\lambda$\cite{ElSawyII} and $X_i$ denotes the location of the BS $i$. A user served by a BS $X_i$ is located uniformly at random within the Voronoi cell of $X_i$. The typical user is located within the \emph{typical cell}, the Voronoi cell of the typical BS at origin. The distance between the typical user and the typical BS of $\Phi$ is $D$. Its approximate distribution is $D\sim$ Rayleigh$\left(\sigma\right)$, with the scale parameter $\sigma=1/\sqrt{2\pi\lambda}$\cite{RHI}. We consider a translated version of the PPP $\Phi$ so that the typical user is at the origin.

On the downlink, each BS transmits a $K$-bit packet to its user using a physical layer rateless code. Each BS transmits with constant power $\rho$. The channel is quasi-static flat fading affected by path loss. The interference power and SIR at the
typical user based on the typical BS transmission are given by
\begin{equation}
I=\sum_{k\neq 0}\rho h_{k} \abs{X_k}^{-\alpha}
\label{int_eq}
\end{equation}
\begin{equation}
\mathrm{SIR}=\frac{\rho h D^{-\alpha}
}{I}\label{sir_in},
\end{equation}
where $h$ and $h_k$ have $Exp(1)$ distribution.

Each packet transmission of $K$ bits has a delay constraint of $N$ channel uses. Define $\hat{T}$ as the time to decode a $K$-bit packet, and $T$ as the packet transmission time. They are defined as
\begin{align}
&\hat{T}\triangleq \min\left\{t:K<t\cdot C\right\}\label{Rx_pkt}\\
&T\triangleq \min (N,\hat{T}),\label{pkt_Ti}
\end{align}
where $C=\log_2\left(1+ \mathrm{SIR}\right)$ is the achievable rate of the typical BS transmission and depends on the type of receiver used. Note that $T$ is a truncated version of $\hat{T}$ at sample value $t=N$.

Now, we focus on a framework introduced in \cite{MeDi_Pap} to study the network performance conditioned on the PPP $\Phi$. In this letter, the two metrics used to quantify the performance of rateless coded transmission are the success probability and the rate of $K$-bit packet transmission conditioned on $\Phi$, defined as
\begin{align}
P_s(N)&\triangleq 1-\mathbb{P}(\hat T> N\mid \Phi)\label{p_s}\\
R_N&\triangleq \frac{KP_s(N)}{\mathbb{E}\left[T\mid \Phi\right]}.\label{Rn}		
\end{align}
Since $T$ is basically $\hat{T}$ truncated at sample value $t=N$, both $P_s(N)$ and $R_N$ depend on the distribution of $\hat{T}$ conditioned on $\Phi$. $R_N$ in (\ref{Rn}) is a random variable (RV). It quantifies the per-user rate achieved in a given PPP realization $\Phi$.

From (\ref{pkt_Ti}), the CCDF of $T$ is
$\mathbb{P}\left(T>t\right)=\mathbb{P}(\hat{T}>t)$, $t<N$. Plugging the expression for $C$ in (\ref{Rx_pkt}), we obtain
\begin{align}
&\mathbb{P}(\hat{T}>t)
=\mathbb{P}\left(K/t\geq \log_2\left(1+ \mathrm{SIR}\right)\right)\label{ccdf_eq}\\
&P_s(t)\triangleq \mathbb{P}(\hat{T}\leq t\mid \Phi)=\mathbb{P}\left(\mathrm{SIR}\geq \theta_t\mid \Phi\right),\label{cdf_con}
\end{align}
where $\theta_t=2^{K/t}-1$. Note that the CDF in (\ref{cdf_con}) is a RV due to conditioning on $\Phi$. Below, we discuss the conditional packet transmission time distribution for two types of interference models. One type is the time-varying interference (TvI) model and the second type is the constant interference (CI) model. 
\section{Theoretical Analysis}
\label{AnaResu}
For the TvI model, we assume that the interfering BSs transmit only a $K$-bit packet to their user and turn off afterwards. Since the packet transmission time of each interfering BS is random, the interference at the typical user is time-varying. The time-averaged interference up to time $t$ is given by
\begin{equation}
\hat{I}(t)=\sum_{k\neq 0}\rho h_{k} \abs{X_k}^{-\alpha}\min\left(1,T_k/t\right),
\label{int_avr}
\end{equation}
where $T_k$ is the packet transmission time of BS $X_k$. Assuming each user employs a nearest-neighbor decoder, the achievable rate at the typical user is given by
\begin{equation}\label{Ct}
C(t)=\log_2\left(1+ \mathrm{SIR}(t)\right),
\end{equation}
where $\mathrm{SIR}(t)$ is obtained from (\ref{sir_in}) with $I$ being replaced by $\hat{I}(t)$ in (\ref{int_avr}). The packet transmission time distribution depends on the Laplace transform (LT) of the interference $\hat{I}(t)$. However, due to the fact that the marks $T_k$ are correlated, it is infeasible find the LT of $\hat{I}(t)$. For the sake of analysis, we consider an approximation termed the independent thinning model (ITM) in which the correlated marks $T_k$ are replaced by i.i.d. marks $\bar{T}_k$ with a given CDF $F(\bar{t})$. The average interference up to time $t$ under ITM is given by
\begin{equation}
\bar{I}(t)=\sum_{k\neq 0}\rho h_{k} \abs{X_k}^{-\alpha}\min\left(1,\bar{T}_k/t\right).
\label{iavITM}
\end{equation}
Let $\bar{\eta}_k(t)=\min\left(1,\bar{T}_k/t\right)$. We just use $\bar{\eta}_k$
for simplicity. The corresponding achievable rate is obtained from (\ref{Ct}) and in the expression for $\mathrm{SIR}(t)$, $\bar{I}(t)$ in (\ref{iavITM}) is used.
The analysis of the TvI model is based on the ITM assumption throughout the paper. 
The CDF of $\hat T$ when conditioned on $\Phi$ under the ITM is given by
\begin{align}
P_s(t)&\triangleq \mathbb{P}(\hat T\leq t\mid \Phi)=\mathbb{P}\Big(\frac{\rho h D^{-\alpha}}{\bar{I}(t)}\geq \theta_t \mid \Phi\Big)\label{BapCP}\\
&=\mathbb{P}\Big(h \geq \theta_t D^{\alpha}\sum_{k\neq 0}h_{k} \abs{X_k}^{-\alpha}\bar{\eta}_k \mid \Phi \Big)\nonumber\\
&=\prod_{k\neq 0}\mathbb{E}_{\bar{\eta}}\left[\frac{1}{1+\theta_t \left(D/\abs{X_k}\right)^{\alpha}\bar{\eta}_k} \right]\label{Exeq}\\
&\stackrel{(a)}{\geq} \prod \frac{1}{1+\theta_t \left(D/\abs{X_k}\right)^{\alpha}\mathbb{E}\left[\bar{\eta}_k\right]}
\label{LBeq},
\end{align}
where $(a)$ follows by applying the Jensen's inequality for convex functions in (\ref{Exeq}). The moments of the RV $P_s(t)$ in (\ref{BapCP}) are bounded below.
\begin{Theoi1}
\label{Th1Mn}
The moments of the conditional CDF of the packet transmission time under the independent thinning model, $P_s(t)$ in (\ref{BapCP}), are bounded as
\begin{align}
M_n&\triangleq\mathbb{E}\left[\left(P_s(t)\right)^n\right]\geq \frac{1}
{{}_2F_{1}\left(\left[n,-\delta\right];1-\delta;-\omega(t)\theta_t\right)}\label{MntvI}\\
\omega(t)&=\frac{1}{t}\int_{0}^{t} 1-F(x) \ud x \label{ome_t}\\
F(t)&=\frac{1}{{}_2F_{1} \left(\left[1,-\delta\right]; 1-\delta; -\theta_t\min\left(1,\mu/t\right)\right)}\label{TkCDF}\\
\mu&=\int_0^{N} \left(1-{}_2F_{1}\left(\left[1,\delta\right]; 1+\delta;-\theta_t\right)\right)\ud t\label{mu_exp},
\end{align}
where ${}_2F_{1}\left([a, b]; c; z\right)$ is the Gauss hypergeometric function and $\delta=2/\alpha$.
\end{Theoi1}
\begin{IEEEproof}
Letting $\bar{\theta}_t=\theta_t\mathbb{E}\left[\bar{\eta}_k\right]$ in (\ref{LBeq}), we get
\begin{align}
&M_n=\mathbb{E}\left[\left(P_s(t)\right)^n\right]
\geq \mathbb{E}\Bigg[\prod \frac{1}{\left(1+\bar{\theta}_t \left(D/\abs{X_k}\right)^{\alpha}\right)^n}\Bigg]\label{Mnb}\\
&\stackrel{(a)}{=}\mathbb{E}\Bigg[\exp\left(-\pi\lambda \int_D^{\infty} \Big(1- \frac{1}{\left(1+\bar{\theta}_t \left(D/v\right)^{\alpha}\right)
^n}\Big)\ud v^2\right)\Bigg]\nonumber
\end{align}
\begin{align}
&\stackrel{(b)}{=}\mathbb{E}\Bigg[\exp \Bigg(-\pi\lambda D^2 \underbrace{\delta \bar{\theta}_t^{\delta}\int_0^{\bar{\theta}_t}\Big(1-\frac{1}{\left(1+y\right)^n}\Big) \frac{\ud y}{y^{1+\delta}}}_{H\left(\bar{\theta}_t\right)} \Bigg)\Bigg]\nonumber\\
&\stackrel{(c)}{=}\frac{1}{1+H\left(\bar{\theta}_t\right)}=\frac{1}{{}_2F_{1}\left( \left[n, -\delta\right];1-\delta;-\theta_t \mathbb{E}\left[\bar{\eta}\right]\right)}\label{fetqn},
\end{align}
where (a) is due to the PGFL of the uniform PPP $\Phi_b$, (b) follows by using $y=\bar{\theta}_t \left(D/v\right)^{\alpha}$ and the $\mathbb{E}[\cdot]$ operation w.r.t $D$ leads to (c). The function $H(\bar{\theta}_t)$ can be expressed as in (\ref{fetqn}).
Define $\omega(t)\triangleq \mathbb{E}\left[\bar{\eta}(t)\right]=\mathbb{E}\left[\min\left(1,\bar{T}/t\right)\right] =\int_0^1\mathbb{P}(\bar{T}/t>x)\ud x$, which depends on the given CDF of $\bar{T}$. Now $F(\bar{t})$, the CDF of $\bar{T}$ is given in (\ref{TkCDF}) and (\ref{mu_exp}).
\end{IEEEproof}
Note that the above function $H(\bar{\theta}_t)$ was based on the lower bound in (\ref{LBeq}). The corresponding function that can be obtained based on the exact expression in (\ref{Exeq}) is given below.
\begin{align}
G\left(\theta_t\right)&=\delta \theta_t^{\delta} \int_0^{\theta_t} \Big(1-\left(\mathbb{E}\left[\frac{1}{1+\bar{\eta} y}\right]\right)^n\Big) \frac{\ud y}{y^{1+\delta}}\label{Mnexa}\\
M_n&=\frac{1}{1+G\left(\theta_t\right)}\geq \frac{1}{1+H\left(\bar{\theta}_t\right)} \triangleq \tilde{M}_n\label{Mnti}.
\end{align}
Note $G\left(\theta_t\right)$ in (\ref{Mnexa}) can be expanded in closed form. However, it is not feasible to provide a simple expression for $G\left(\theta_t\right)$ in terms of the hypergeometric function similar to $H(\bar{\theta}_t)$. Hence, we work with the bound $H(\bar{\theta}_t)$ for tractability and use the bound $\tilde{M}_n$ defined in (\ref{Mnti}) instead of $M_n$. Below, we derive two key results founded on Theorem \ref{Th1Mn}. Under the CI model, it is assumed that the interfering BSs are transmitting \emph{continuously} to their users for the entire duration of the typical user reception time. The typical BS transmits a $K$-bit packet to the typical user and the performance under such a scenario is quantified below. The CI model is a special case of the TvI model. We quantify the moments of the RV $P_s(t)$ in (\ref{cdf_con}).

\begin{Coro1}
\label{Th1ci}
The moments of the conditional CDF of the packet transmission time under the constant interference model, $P_s(t)$ in (\ref{cdf_con}), are given by
\begin{equation}\label{MnCI}
M_n\triangleq\mathbb{E}\left[\left(P_s(t)\right)^n\right]=\frac{1}
{{}_2F_{1}\left(\left[n,-\delta\right];1-\delta;-\theta_t\right)}.
\end{equation}
\end{Coro1}
\begin{IEEEproof}
For the CI model, $\bar{\eta}_k=1$. Hence, plugging $\mathbb{E}\left[\bar{\eta}\right]=1$ in (\ref{fetqn}) yields the desired result in (\ref{MnCI}).
\end{IEEEproof}
Now, we focus on the two RVs $P_s(t)$ and $P_s(u)$, $t\neq u$ for the TvI model and quantify their product moment.
\begin{Theoi2}
\label{Th2Mn}
The product moment of the two RVs $P_s(t)$ and $P_s(u)$, $t\neq u$ as defined in (\ref{BapCP}) is given by
\begin{align}
&\mathbb{E}\left[ P_s(t) P_s(u) \right]\geq \frac{1} {1+J\left(\bar{\theta}_t,\bar{\theta}_u\right)} \label{Eptpu}\\
&J\left(\bar{\theta}_t,\bar{\theta}_u\right)=\delta \int_0^1\Big[1-\frac{1}{\left(1+\bar{\theta}_t y\right)\left(1+\bar{\theta}_u y\right)}\Big] \frac{\ud y}{y^{1+\delta}},\label{J2exp}
\end{align}
where $\bar{\theta}_t=\omega(t) \theta_t$, $\bar{\theta}_u=\omega(u) \theta_u$ and $\omega(t)$ is defined in (\ref{ome_t}).
\end{Theoi2}
\begin{IEEEproof}
The result in (\ref{Eptpu}) and (\ref{J2exp}) is derived using steps similar to that in the proof of Theorem \ref{Th1Mn}. Due to space constraints, we only highlight the differences. The $P_s(t)$ and $P_s(u)$ expressions are given in (\ref{Exeq}) and (\ref{LBeq}). Hence, we obtain
\begin{equation}
P_s(t)P_s(u)\geq \prod \frac{1}{\left(1+\bar{\theta}_t \left(D/\abs{X_k}\right)^{\alpha}\right)\left(1+\bar{\theta}_u \left(D/\abs{X_k}\right)^{\alpha}\right)}\label{Pstu}.
\end{equation}
Define $Q\triangleq \mathbb{E}\left[ P_s(t) P_s(u) \right]$, where $\mathbb{E}\left[\cdot\right]$ is w.r.t $\Phi$. Now based on (\ref{Pstu}), $Q$ is computed using the PGFL of $\Phi$ with steps similar to (\ref{Mnb})-(\ref{fetqn}), except that the function $H\left(\bar{\theta}_t\right)$ in Theorem \ref{Th1Mn} is replaced by $J\left(\bar{\theta}_t,\bar{\theta}_u\right)$ defined in (\ref{J2exp}) respectively.
\end{IEEEproof}
\begin{equation}\label{Qftn}
Q= \mathbb{E}\left[ P_s(t) P_s(u) \right]\geq \frac{1} {1+J\left(\bar{\theta}_t,\bar{\theta}_u\right)}\triangleq \tilde{Q}.
\end{equation}
Similar to Theorem \ref{Th1Mn}, we work with the bound $J\left(\bar{\theta}_t,\bar{\theta}_u\right)$ for tractability and use the bound $\tilde{Q}$ defined in (\ref{Qftn}) from now onwards instead of $Q$. 

\section{Distribution Approximations}
\label{dis_appr}
Based on the moments $\tilde{M}_n$, a closed form expression for the CDF (or its bounds) of the RV $P_s(t)$ in (\ref{BapCP}) can be written. Since $P_s(t)$ is supported on the interval [0,1], the beta distribution will yield a simple yet useful alternative. We also obtain a distribution approximation for $R_N$ in (\ref{Rn}).
\subsection{Beta Approximation of $P_s(t)$ for TvI and CI models}
\label{betapr}
The PDF of beta-approximated $P_s(t)$ is given by
\begin{equation}\label{be_pdf}
f(x)=\frac{x^{\bar \gamma-1}\left(1-x\right)^{\beta-1}}{\mathrm{B}\left(\bar \gamma,\beta\right)},~~x \in [0,1],
\end{equation}
where $B\left(a, b\right)=\int_{0}^{1} x^{a-1} (1-x)^{b-1} \ud x$ is the beta function, $\bar \gamma$ and $\beta$ are related to the moments of $P_s(t)$ as \cite{MeDi_Pap}
\begin{align}
&\bar \gamma = \frac{\gamma_1 \beta}{1-\gamma_1}; ~~\beta=\frac{(\gamma_1-\gamma_2)(1-\gamma_1)}{\gamma_2-\gamma_1^2}, \label{bepa}
\end{align}
where $\gamma_n=\tilde{M}_n$. Both parameters $\bar \gamma$ and $\beta$ are functions of $t$. Note for the CI model, the moments $\tilde{M}_n$ are given in (\ref{MnCI}). Now we provide the CCDF result for $P_s(t)$ at $t=N$.

\begin{Propi1}
\label{PrPsn}
The CCDF of the per-user coverage probability $P_s(N)$ in (\ref{p_s}) is approximated as
\begin{align}
\mathbb{P}(P_s(N)>p)&=\frac{\bar{\mathrm{B}}\left(p, \bar \gamma, \beta\right)}{\mathrm{B}\left( \bar \gamma, \beta\right)} \label{Ps_dis},
\end{align}
where $\bar{\mathrm{B}}\left(a, b, c\right)=\int_{a}^{1} y^{b-1}\left(1-y\right)^{c-1} \ud y$ is the upper incomplete beta function and $\bar \gamma$, $\beta$ are defined in (\ref{bepa}).
\end{Propi1}
\begin{IEEEproof}
Using the beta approximation for $P_s(N)$ in (\ref{be_pdf}) completes the proof.
\end{IEEEproof}

\subsection{$R_N$ Distribution Approximation for TvI and CI models}
\label{PsRn}
From (\ref{Rn}), the distribution of $R_N$ can be obtained from the PDF of $P_s(N)$ given in (\ref{be_pdf}) and (\ref{bepa}) with $t=N$ and also, the distribution of $\mathbb{E}\left[T\mid \Phi\right]$. Let
\begin{equation}\label{etcon}
T_\phi\triangleq \mathbb{E}\left[T\mid \Phi\right]=\int_0^N \left(1-\mathbb{P}(\hat T\leq t\mid \Phi)\right) \ud t.
\end{equation}

Now, the CCDF of rate $R_N$ in (\ref{Rn}) is given by
\begin{align}
\mathbb{P}(R_N>r)&=\mathbb{E}\left[\mathbb{P}\left(P_s(N)>\frac{rT_\phi}{K}\Big | T_\phi\right)\right]\nonumber\\
&=\frac{\mathbb{E}\left[\bar{\mathrm{B}}\left(rT_\phi/K, \bar \gamma, \beta\right)\right]}{\mathrm{B}\left( \bar \gamma, \beta\right)},\label{Rnpdf}
\end{align}
To evaluate (\ref{Rnpdf}), the distribution of $T_\phi$ in (\ref{etcon}) is very critical. We first obtain the moments of $T_\phi$.

From (\ref{etcon}), the first two moments of $T_\phi$ are given by
\begin{align}
\nu_1&= \mathbb{E}\left[T_\phi\right]=N-\int_0^N \tilde{M}_1(t)\ud t \label{TpE1}\\
\nu_2&=\mathbb{E}\left[T_\phi^2\right]=\mathbb{E}\Big[\big(N-\int_0^N P_s(t)\ud t\big)^2\Big]\nonumber\\
&\stackrel{(a)}{=} N\left(2\nu_1-N\right)+ \mathbb{E}\Big[\big(\int_0^N P_s(t)\ud t\big)^2\Big],\label{TpE2}
\end{align}
where in (a), the second term is given below.
\begin{align}
\mathbb{E}\Big[\big(\int_0^N P_s(t)\ud t\big)^2\Big]&=
\mathbb{E}\Big[\int_0^N P_s(t)\ud t \int_0^N P_s(u)\ud u\Big]\nonumber\\
&=\int_0^N \int_0^N \mathbb{E}\Big[ P_s(t) P_s(u) \Big] \ud t \ud u\nonumber\\
&\stackrel{(b)}{=}\int_0^N \int_0^N \frac{1} {1+J\left(\bar{\theta}_t,\bar{\theta}_u\right)} \ud t \ud u \label{Ept2},
\end{align}
where (b) is obtained from Theorem \ref{Th2Mn}.

Using (\ref{Ept2}) and (\ref{J2exp}), the second moment $\nu_2$ in (\ref{TpE2}) can be computed. For the CI model, the moment expressions are exact. However for the TvI model, the moment expressions are bounds since they are based on the tractable bound of $P_s(t)$ in (\ref{LBeq}). It is not feasible to express $T_\phi$ moments in closed form based on the exact $P_s(t)$ in (\ref{Exeq}). The PDF of $T_\phi \in [0,N]$ is approximated by a known distribution, whose parameters are expressed in terms of the moments of $T_\phi$.

Beta distribution has been used widely to approximate the distribution of a RV with finite support. Hence, we propose to model $T_\phi/N$ as a beta distributed RV. The first two moments of $T_\phi/N$ are given by
\begin{equation}\label{TNmom}
\kappa_1=\frac{\nu_1}{N};~~\kappa_2=\frac{\nu_2}{N^2}.
\end{equation}
Now, the two parameters $\bar{\kappa}$ and $\vartheta$ of the beta distribution for the RV $T_\phi/N$ are given by
\begin{equation}\label{betpar}
\bar{\kappa}=\frac{\kappa_1\vartheta}{1-\kappa_1};~~ \vartheta=\frac{(\kappa_1-\kappa_2)(1-\kappa_1)}{\kappa_2-\kappa_1^2}.
\end{equation}

The PDF of $T_\phi/N$ is similar in form to (\ref{be_pdf}) except for the parameters $\bar{\kappa}$ and $\vartheta$. Below, we summarize the main result.

\begin{Theoi3}
\label{Th3Rn}
The CCDF of the per-user rate $R_N$ in (\ref{Rn}) is approximated as
\begin{align}
\mathbb{P}(R_N>r)&=\int_{0}^{1}\frac{\bar{\mathrm{B}}\left(rNy/K, \bar \gamma, \beta\right)}{\mathrm{B}\left( \bar \gamma, \beta\right)} \frac{y^{\bar{\kappa}-1}(1-y)^{\vartheta-1}}{\mathrm{B}\left(\bar{\kappa}, \vartheta\right)} \ud y.\label{Rn_dis}
\end{align}
\end{Theoi3}
\begin{IEEEproof}
The CCDF of $R_N$ is given in (\ref{Rnpdf}). Using the beta approximation for $T_\phi/N$ in (\ref{Rnpdf}) completes the proof.
The parameters used in (\ref{Rn_dis}) are defined in (\ref{bepa}) and (\ref{TNmom})-(\ref{betpar}). The moments $\nu_i$ are given in (\ref{TpE1})-(\ref{Ept2}).
\end{IEEEproof}
The CCDF result in (\ref{Rn_dis}) is used in the section on numerical results. Note that the Theorem \ref{Th3Rn} applies to both the TvI and CI models. The result in (\ref{Rn_dis}) is based on modeling $T_\phi/N$ as a beta distributed RV. The time to decode a $K$-bit packet $\hat{T}$ in (\ref{Rx_pkt}) has been fitted with a Gamma distribution in \cite{RHI}. Since $\hat{T}\in [0,\infty)$, it seems Gamma PDF is a better match. However, $T_\phi \in [0,N]$ and hence, beta PDF is justified for $T_\phi/N$.
\subsection{Fixed-Rate Coding}
\label{frc}
For fixed-rate coding, the per-user rate is defined as
\begin{equation}
R_N\triangleq \frac{K}{N}~\mathbb{P}\left(\mathrm{SIR}>2^{K/N}-1\mid \Phi\right). \label{FRC_CP}
\end{equation}
The CCDF of the rate $R_N$ is given by
\begin{equation}\label{Frc_rn}
\mathbb{P}(R_N>r)=\frac{\bar{\mathrm{B}}\left(rN/K, \bar \gamma, \beta\right)}{\mathrm{B}\left( \bar \gamma, \beta\right)}.
\end{equation}
In the expression for parameters $\bar \gamma$ and $\beta$ in (\ref{Frc_rn}), the moments $M_n$ given in (\ref{MnCI}) are used.
Adaptive modulation and coding (AMC) is a scheme used in current 4G networks to adapt the rate to channel conditions\cite{GoldsmithII,LTEBook,CaireII}. AMC chooses a fixed-rate code and its code rate to most closely match the channel conditions. The analytical discussions in the current paper focused on rateless codes are also applicable to the case of AMC with the important change of packet time $t$ resolution. For the rateless case, the packet time $t\in \mathbb{N}$ whereas for the AMC case, the resolution changes to $t=\{N_i\}$, where $N_i$ is the number of parity symbols for the AMC index $i$.

The CCDF of the rate $R_N$ with fixed-rate coding and power control is also given by (\ref{Frc_rn}). However to compute the parameters $\bar \gamma$ and $\beta$, the moments $M_n$ are obtained from \cite{Wang}.

\subsection{Insights from Theorem \ref{Th3Rn}}
\label{The3Insi}
\begin{itemize}
  \item From (\ref{Frc_rn}), we can see that the tail of the $R_N$ CCDF for fixed-rate coding decays very rapidly as $r\rightarrow K/N$. However for rateless coding, it can be observed from (\ref{Rnpdf}) and (\ref{Rn_dis}) that the decay of the tail is much slower. The tail of the rateless $R_N$ CCDF is more heavy tail-like and has a slower decay due to the $\mathbb{E}[\cdot]$ w.r.t $T_\phi/N$.
  \item Presence of a heavy tail for the $R_N$ CCDF implies a smaller total energy consumption for a $K$-bit packet transmission, thus resulting in enhanced energy-efficiency in the cellular downlink.

  \item In the $R_N$ expressions of (\ref{Rn}) and (\ref{FRC_CP}), we can see that both the schemes have the $P_s(N)$ term. The distribution of $P_s(N)$ is the same for both fixed-rate coding and rateless coding under the CI model. The profound impact in $R_N$ distribution for the rateless case arises due to the $T_\phi$ term, which is a result of variable-length coding in the physical layer leading to a variable transmission time.
\end{itemize}
\section{Numerical Results}
\label{sec:Num_Res}
\begin{figure}[!hbtp]
\centering
\includegraphics[scale=0.55, width=0.5\textwidth]{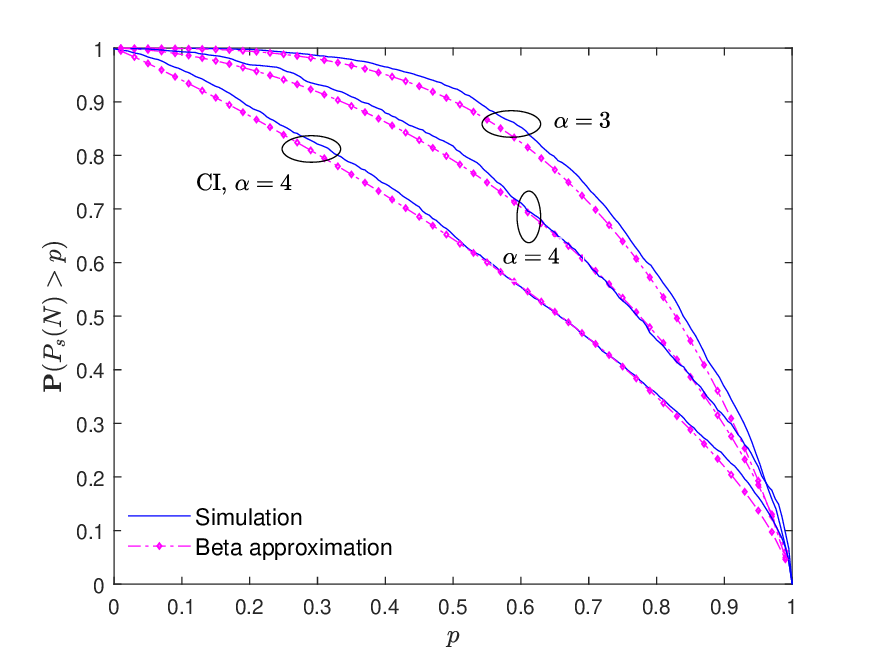}
\caption{The CCDF of the per-user coverage probability $P_s(N)$ in (\ref{p_s}) in a cellular downlink with $\lambda=1$, $\alpha=\{3,4\}$ and $N=\{200, 90\}$ respectively. The analytical curve is based on (\ref{Ps_dis}).}
\label{Psccdf}
\end{figure}

In this section, numerical results showing the efficacy of the proposed per-user performance analysis are presented. For the network simulation, the following parameters were chosen: $\lambda=1$ and $K=75$.
Fig. \ref{Psccdf} shows the plots of CCDF of $P_s(N)$ for the rateless coding scenario. It is observed that the curves corresponding to the beta distribution approximation for $P_s(N)$ given in Proposition 1 matches the simulation curves very well for both the TvI and CI models.
\begin{figure}[!hbtp]
\centering
\includegraphics[scale=0.55, width=0.5\textwidth]{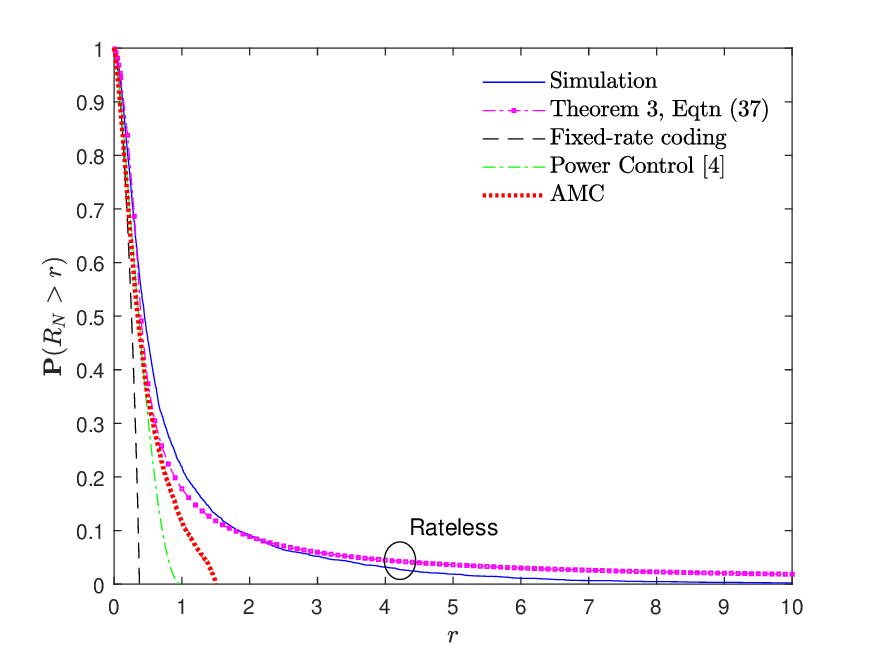}
\caption{The CCDF of the per-user rate $R_N$ in (\ref{Rn}) in a cellular downlink with $\lambda=1$, $\alpha=3$ and $N=200$. The rateless coding curve is based on (\ref{Rn_dis}), while the fixed-rate coding curves for both constant power and power control are based on (\ref{Frc_rn}). For rateless coding, the curve is based on the \emph{constant interference } (CI) model.}
\label{Rnccdf}
\end{figure}
\begin{figure}[!hbtp]
\centering
\includegraphics[scale=0.55, width=0.5\textwidth]{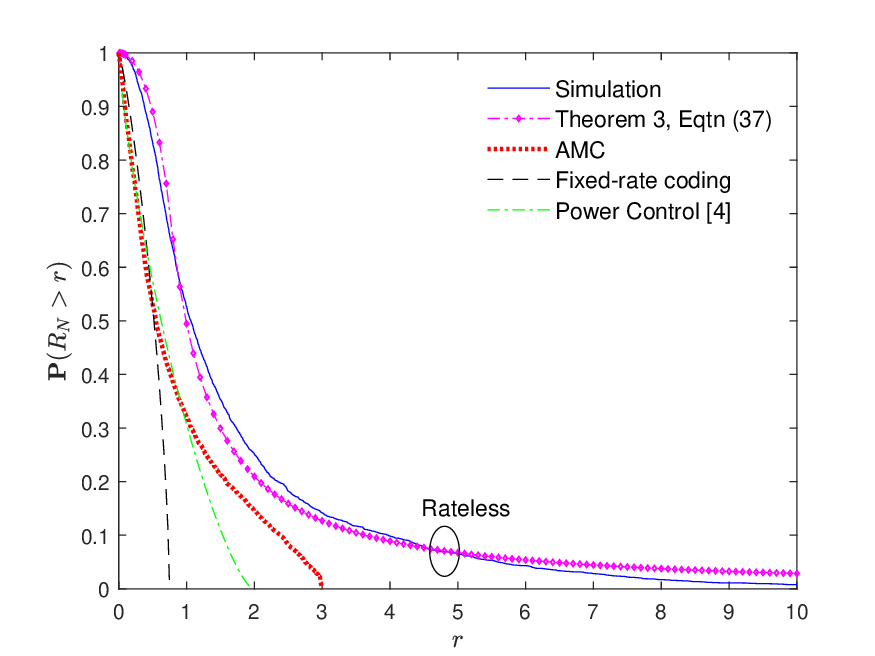}
\caption{The CCDF of the per-user rate $R_N$ in (\ref{Rn}) in a cellular downlink with $\lambda=1$, $\alpha=4$ and $N=100$. For rateless coding, the curve is based on the \emph{time-varying interference} (TvI) model.}
\label{Rnccdftvial4}
\end{figure}

In Fig. \ref{Rnccdf}, a plot of the CCDF of the per-user rate $R_N$ in a cellular network at $\alpha=3$ and $N=200$ is shown. Fig. \ref{Rnccdftvial4} shows a plot of the CCDF of the per-user rate $R_N$ for a cellular downlink at $\alpha=4$ and $N=100$.
The focus is on comparing three types of transmission schemes, i.e., fixed-rate coding based on (\ref{Frc_rn}), AMC as described in Section \ref{frc} and the rateless coding scheme as per (\ref{Rn_dis}). Note that the performance of AMC is based on simulation only. For the AMC case, we consider a list of four AMC indices. The packet time for AMC index $i$ is set to $t=i\cdot N/4$, $1\leq i \leq 4$. As mentioned before in Section \ref{frc}, the rate $R_N$ for AMC case can be computed as per (\ref{Rn}) taking into consideration the above defined AMC packet times.

In terms of matching the rate to the instantaneous channel conditions, fixed-rate coding with constant power has poor efficiency and rateless coding has high efficiency while AMC and fixed-rate coding with power control have intermediate performance. High efficiency of rateless coding is captured by the term $\mathbb{E}\left[T\mid \Phi\right]$ in the expression for $R_N$ in (\ref{Rn}). For fixed- rate coding, the packet time is fixed to $N$. AMC is also based on fixed-rate codes and thus, the packet time does not change with a finer resolution as compared to rateless coding.

In Fig. \ref{Rnccdftvial4}, the CCDF curve for AMC decays to zero at $r=3$. For rateless coding, the CCDF at $r=3$ is $0.15$. Since rateless codes have robust adaptivity to the instantaneous channel conditions, the scheme yields much higher per-user rates relative to the AMC. These higher per-user rates for the rateless scheme have implications on the energy-efficiency of the BS-UE links and also, the congestion, QoS and end-to-end delay in the network.

Note that the analytical results are based on ITM in Section \ref{AnaResu} and the moment lower bound $\tilde{M}_n$ in (\ref{Mnti}). These two approximations are necessary to obtain simplified expressions. The accuracy of the analytical curves is also influenced by the distribution of downlink distance $D\sim$ Rayleigh$(1/\sqrt{2\pi c\lambda})$. A value of $c=1.25$ has been used in \cite{Wang}. However, in this letter, we use $c=1$ to remain consistent with the majority of the literature\cite{ElSawyII}.

\section{Conclusion}
\label{sec:Concl}
In this letter, we characterize the per-user performance of cellular downlink when physical layer rateless codes are used for adaptive transmission. The BS locations are modeled by a uniform Poisson point process. The performance of rateless codes was presented under both the constant and time-varying interference models. Accurate approximations to the distribution of the per-user coverage probability and transmission rate are derived. The advantages of physical layer rateless codes are clearly illustrated by comparing their performance to fixed-rate based adaptive modulation and coding.

\bibliography{References_SRA}
\bibliographystyle{IEEEtran}
\end{document}